\documentclass[fleqn,usenatbib]{mnras}

\usepackage{newtxtext,newtxmath}

\usepackage[T1]{fontenc}

\DeclareRobustCommand{\VAN}[3]{#2}
\let\VANthebibliography\thebibliography
\def\thebibliography{\DeclareRobustCommand{\VAN}[3]{##3}\VANthebibliography}

\usepackage{graphicx}	
\usepackage{amsmath}	

\title[AGN feedback duty cycle in Planck SZ selected clusters]{AGN feedback duty cycle in Planck SZ selected clusters using \textit{Chandra} observations}

\author[V. Olivares et al.]{
V. Olivares$^{1}$,
Y. Su$^{1}$, P. Nulsen$^{2,3}$, R. Kraft$^{2}$, T. Somboonpanyakul$^{4}$, F. Andrade-Santos$^{2}$, C.~Jones$^{2}$, W.~Forman$^{2}$
\\
$^{1}$ Department of Physics and Astronomy, University of Kentucky, 505 Rose Street, Lexington, KY 40506, USA\\
$^{2}$ Harvard-Smithsonian Center for Astrophysics, 60 Garden St., Cambridge, MA 02138, USA\\
$^{3}$ ICRAR, University of Western Australia, 35 Stirling Hwy, Crawley, WA 6009, Australia
\\
$^{3}$ Kavli Institute for Particle Astrophysics \& Cosmology, P.O. Box 2450, Stanford University, Stanford, CA 94305, USA
}
\date{Accepted XXX. Received YYY; in original form ZZZ}
\pubyear{2022}
\begin{document}
\label{firstpage}
\maketitle
\vspace{-5cm}
\begin{abstract}
We present a systematic study of X-ray cavities using archival \textit{Chandra} observations of nearby galaxy clusters selected by their Sunyaev-Zel'dovich (SZ) signature in the Planck survey, which provides a nearly unbiased mass-selected sample to explore the entire AGN feedback duty cycle. Based on X-ray image analysis, we report that 30 of the 164 clusters show X-ray cavities, which corresponds to a detection fraction of 18\%. After correcting for spatial resolution to match the high-$z$ SPT–SZ sample, the detection fraction decreases to 9\%, consistent with the high-z sample, hinting that the AGN feedback has not evolved across almost 8~Gyrs. Our finding agrees with the lack of evolution of cool-core clusters fraction. 
We calculate the cavity power, $P_{\rm cav}$, and find that most systems of our sample have enough AGN heating to offset the radiative losses of the intracluster medium.

\end{abstract}
\section{Introduction} \label{sec:intro}
\defcitealias{Hlavacek-Larrondo15}{HL15}
\defcitealias{andrade-santos17}{AS17}

Feedback from active galactic nuclei (AGN) jets has been proposed to solve the cooling flow problem (see \citealt{fabian94,McNamara_2005}, for a review). Although the details of how AGN feedback counteracts the radiative losses of the intracluster medium (ICM) in clusters are still not fully understood. Early observations with the X-ray satellite \textit{ROSAT} revealed surface brightness deficits that appear to be spatially aligned with regions of radio emission in the ICM of a few galaxy clusters \citep{boehringer93,carilli94}. Nowadays, with the superb resolution of the X-ray \textit{Chandra} observatory, it has become clear that central AGN located in the Brightest Cluster Galaxy (hereafter BCG) continuously interacts with the surrounding ICM, producing, not solely, the X-ray surface brightness depressions known as X-ray cavities (or bubbles), but also shocks and ripples \citep[e.g.,][]{fabian06}. In addition, high-resolution radio observations by the Jansky Very Large Array (JVLA) have shown that extended radio lobes inflated by the central AGN may excavate these X-ray cavities by pushing aside the surrounding hot gas. Accordingly, they are expected to be filled with radio emission \citep[e.g.,][]{birzan20}. There have also been detections of the so-called ``ghost cavities'' at low radio frequency, which are believed to trace a past AGN outburst, for which the radio emission has faded away. More importantly, the X-ray cavities and bubbles may provide a direct measurement of the work done by the radio-mode feedback on the ICM \citep[e.g.,][]{gitti10}. The X-ray cavities are not only supposed to carry enough energy to balance the cooling losses of the X-ray emitting plasma \citep{birzan08}, but also play a key role in the formation of extended multiphase filaments observed in cooling flow clusters \citep[e.g.,][]{olivares19,2022arXiv220107838O,russell19}. Therefore, investigating the physical properties of X-ray cavities can improve our understanding of AGN feedback and its impact on galaxy formation and evolution.

Currently, most X-ray cavity studies of clusters, groups and elliptical galaxies are based on \textit{Chandra} X-ray observations for both individual systems and dedicated surveys (see \citealt{birzan04,birzan08, rafferty06,nulsen09,dong10,osullivan11,Hlavacek_Larrondo_2013,shin16,panagoulia14b,birzan17}). One of the main limitations of existing studies based on X-ray selection methods is that they are often biased towards bright cool-core systems and, consequently, against X-ray faint clusters. A complete unbiased sample of galaxy clusters is desirable to obtain heating and cooling balance constraints \citep{gitti10}, and to understand the duty cycle of AGN feedback, which is estimated as the fraction of systems displaying bubbles inflated by the central AGN.

Millimeter-wave surveys utilizing the SZ effect have the advantage of providing nearly mass-limited samples, as the impact of SZ effect on the CMB (cosmic microwave background) brightness temperature is independent of redshift. This allows us to explore the entire AGN feedback duty cycle. Examples of instruments used to undertake SZ surveys include the Planck satellite \citep{planck_collaboration11}, the South Pole Telescope (SPT) \citep{bleem15,bleem20}, and the Atacama Cosmology Telescope \citep{hincks10,hilton18}. For the high-$z$ Universe, \citet[][hereafter HL15]{Hlavacek-Larrondo15} performed a \textit{Chandra} study of 83 clusters selected from the SPT-SZ survey, and found X-ray surface brightness depression in 6 clusters consistent with radio jets inflating X-ray cavities in their ICM. Here, we present a study of X-ray cavities in the Planck SZ survey, which provides a unique and unbiased view of AGN feedback in the nearby ($z<0.35$) Universe and anchors the evolution of AGN feedback over the past 8 Gyrs.


This paper examines \textit{Chandra} observations of 164 Planck SZ clusters with the aim of identifying X-ray bubbles. In section~\ref{sec:sample} we describe the Planck SZ sample. Section~\ref{sec:obs} presents the X-ray \textit{Chandra} observations and describes the methods used to identify X-ray surface brightness depressions. Section~\ref{sec:results} is devoted to the results and their implications. Section~\ref{sec:limitations} presents the limitations of the present study. Finally, section~\ref{sec:conclusions} summarizes our findings.

Throughout this paper, we adopted a standard cosmology with H$_{\rm 0}$=70\,km s$^{-1}$\,Mpc$^{-1}$ and $\Omega_{\rm m}$=0.3.

\section{Sample}\label{sec:sample}
The \textit{Chandra}-Planck Legacy Program for Massive Clusters of Galaxies \citep{jones12} is a deep X-ray survey of massive Planck  clusters with redshift $\leq0.35$ detected over almost the full sky (and $|$b$|> 15 \deg$) through the Sunyaev-Zel'dovich effect by the first Planck mission released in early 2011 \citep{planck_collaboration11}. The observations are constructed by combining the \textit{Chandra} XVP (PI: Jones) and HRC Guaranteed Time Observations (PI: Murray). At least 10,000 source counts have been collected for each cluster to derive its gas properties out to $R_{\rm 500}$  \citep[][hereafter AS17]{andrade-santos17}. The \textit{Chandra} Planck sample is nearly an unbiased, mass-selected sample, covering the mass range $7 \times 10^{13} {\rm M}_\odot \le M_{500} \le 2 \times 10^{15} {\rm M}_\odot$. 
The sample consists of 164 clusters, of which a small fraction contain pronounced substructures (35, subclusters), visually identified in the X-ray images \citepalias{andrade-santos17}. Central density is the best known proxy for the central cooling time \citep[e.g.,][]{Su2020}, and has been widely used to classify CC and NCC clusters \citep[e.g.,][]{ruppin21, andrade-santos17}. Based on the central density classification of $n_{\rm core} = 1.5\times$10$^{-2}$~cm$^{-3}$, as presented in  \citetalias{andrade-santos17}, 63 clusters are classified as CC and 101 as NCC clusters. Deprojected temperature and density profiles of clusters in this sample are taken from  \citetalias{andrade-santos17}, to which we refer the reader for a detailed description.

\section{Observation and Analysis}\label{sec:obs}

For each cluster, we used all available \textit{Chandra} observations, including both CCDs ACIS-I and ACIS-S. The data reduction and calibration of \textit{Chandra} observations were carried out using \textit{Chandra} Interactive Analysis of Observations software (CIAO) 4.12, and Chandra Calibration Database (CALDB) 4.9.2.1. The observations were reprocessed using the {\tt chandra\_repro} tool of CIAO. Standard blank sky background files and readout artifacts were subtracted. Point sources were detected in the 0.5-8.0 keV energy band, then masked before performing the spectral and imagining analysis of the clusters. Exposure corrected images were produced in the 0.5–2.0~keV band energy, and used for the X-ray cavity analysis.

Unsharp masked images were produced to help identify X-ray cavities using CIAO tool {\tt aconvolve}. The original image was smoothed twice using a small- and large-scale Gaussian kernel. The highly smoothed image was then subtracted from the less smoothed image, enhancing the inhomogeneities in the residual image. We tried different smoothing lengths for the more heavily smoothed images based on the large scale of the cluster emission, starting on 10 up to 60~kpc. For the less smoothed images, we tried smoothing lengths comparable to the physical size of a cavity, from 1 up to 20~kpc \citep[e.g.][]{rafferty06}. We also examined the residual image after subtracting an elliptical double beta model, which was obtained by fitting a slightly smoothed 0.5--2.0 keV image. The second beta model is to account for excess emission from the cool core.

We classified each cavity identified as Certain (C) or Potential (P). The first two co-author independently looked for X-ray cavities and then classified them based on the significance of each cavity.

\begin{figure}
\centering
    \includegraphics[width=0.82\columnwidth]{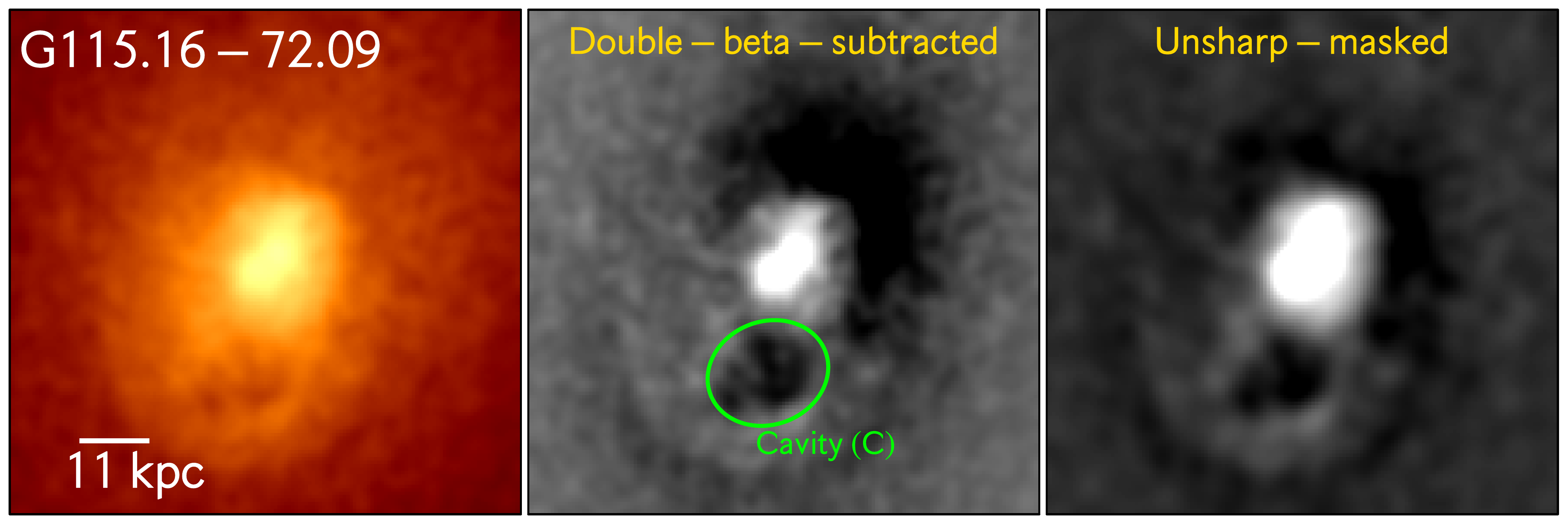}\\
    \vspace{-0.05cm}
	\includegraphics[width=0.82\columnwidth]{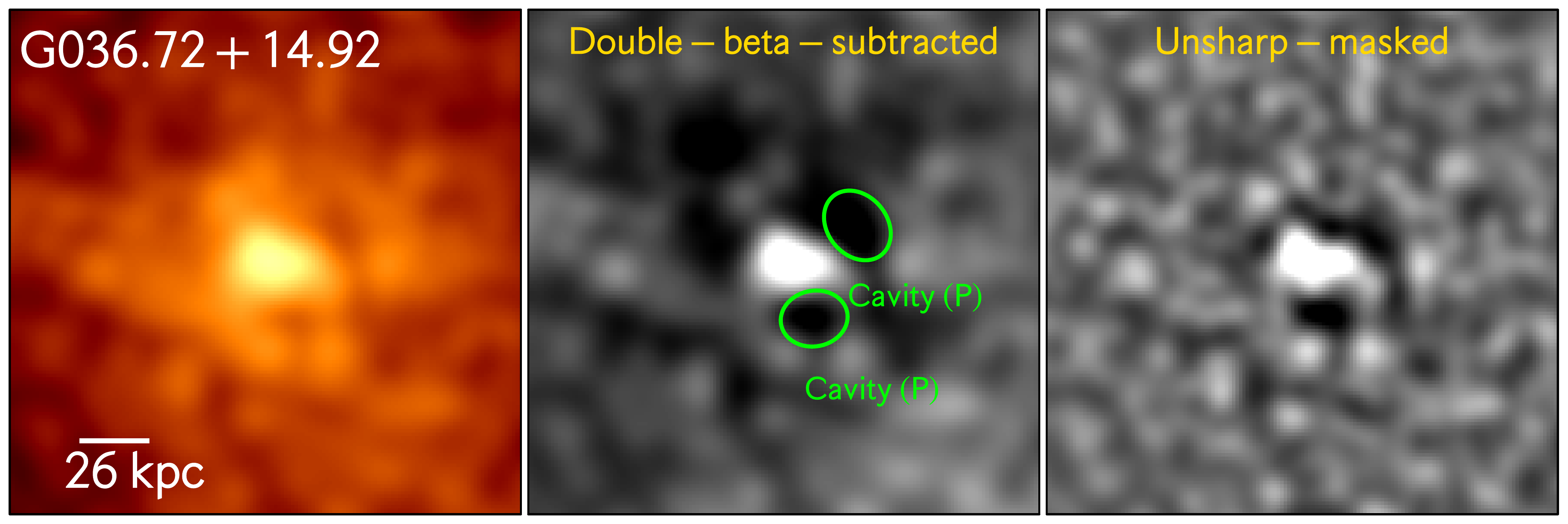}\\
    \vspace{-0.05cm}
	\includegraphics[width=0.82\columnwidth]{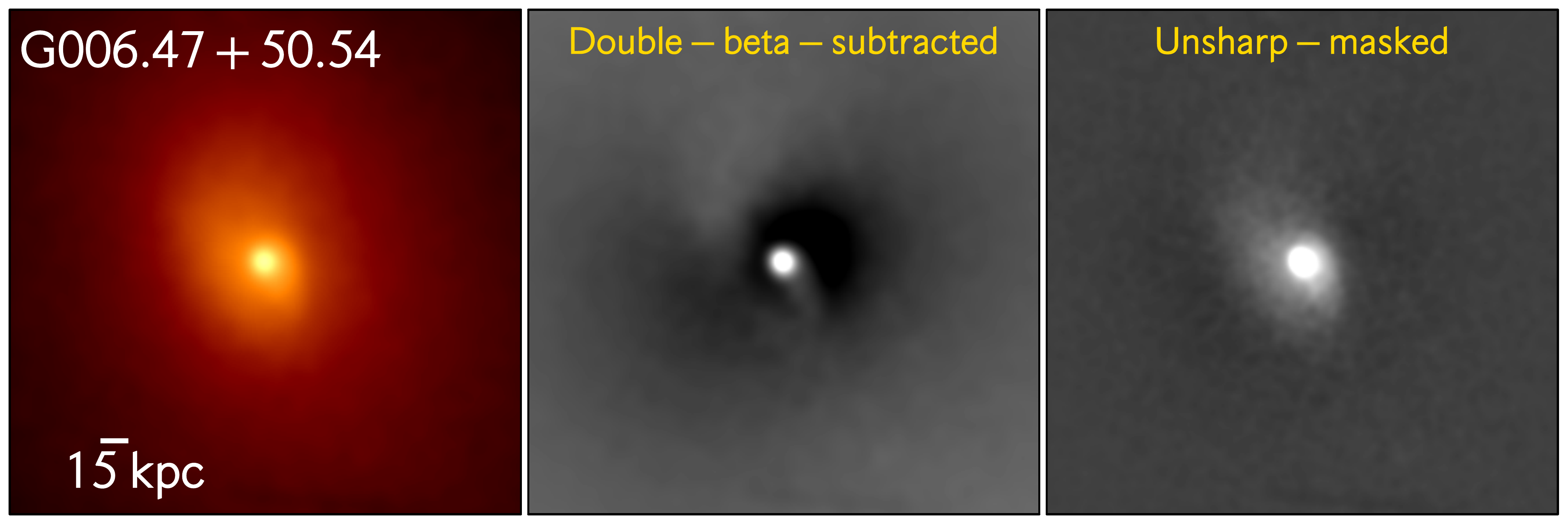}\\
    \caption{Example of the methods employed to look for X-ray cavities for a  cluster with Certain (top), Potential (middle) detected cavities, and without (bottom) cavities. From left to right: 0.5-2.0~keV original image, double $\beta$-subtracted image, and unsharp image. Detected cavities are displayed with green ellipses. \label{fig:example}}
\end{figure}

Cavities were classified as certain if they appear as a clear visible depression in the original image, but also the unsharp-masked image or double $\beta$-subtracted image. In figure~\ref{fig:example} we present an example of the methods employed to identify cavities for a clusters with certain (C), potential (P), and without cavities. A cavity was classified as potential if there was only a hint of an X-ray depression in the original X-ray image, but visible in the unsharp-masked or double $\beta$-model subtracted image. The number of counts of the central region ($<$20~kpc) of the clusters with potential cavities is too low for the cavities to be certain (see also Section~\ref{sec:limitations}). Clusters without depressions were classified as lacking cavities. We also consider clusters with dark annuli or rings created by bright excesses and asymmetries of the cluster distribution as lacking cavities, as such surface brightness depressions are not consistent with bubbles inflated by radio jets. 

\begin{figure*}
    \includegraphics[width=0.82\columnwidth]{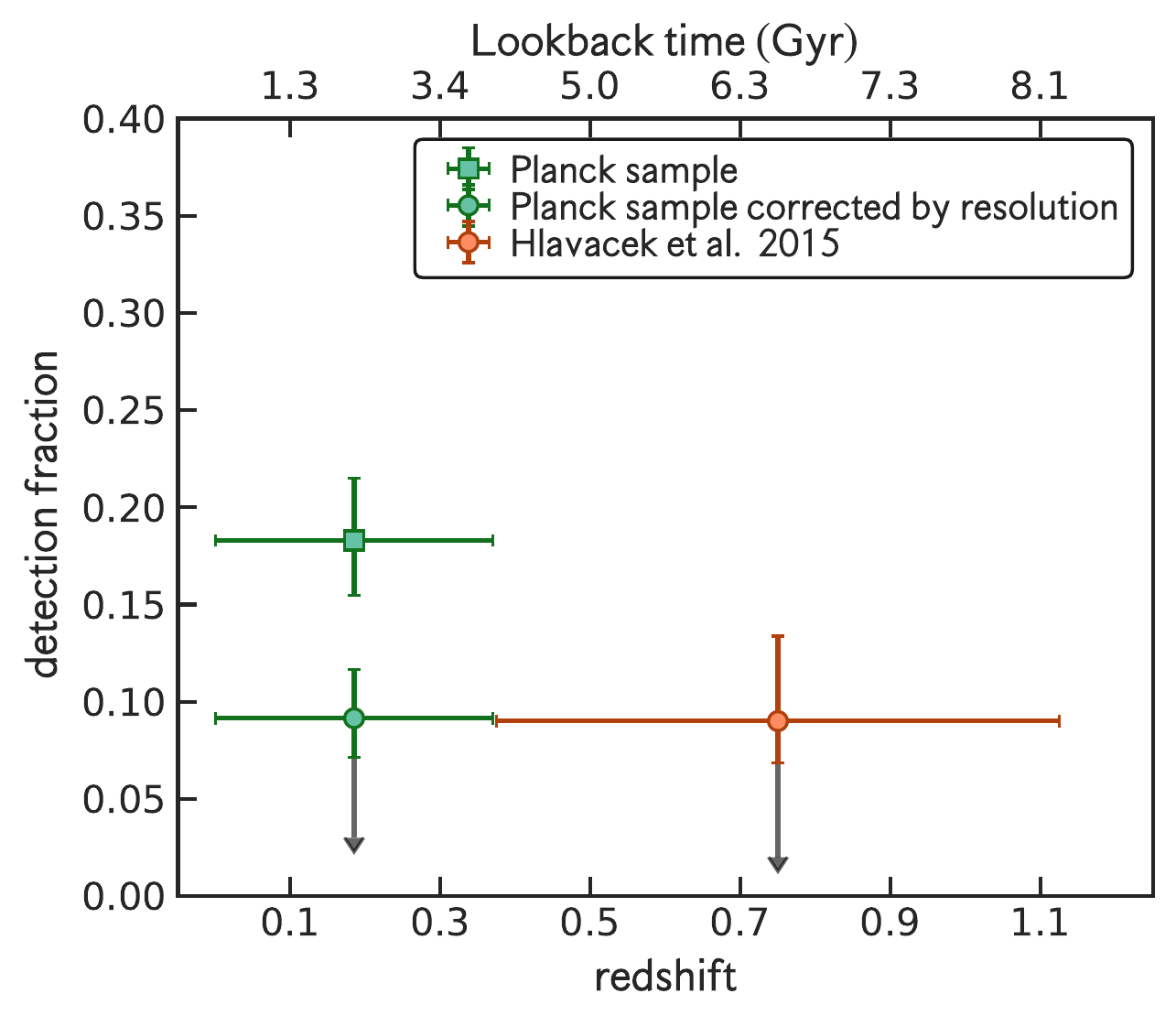}
	\includegraphics[width=0.87\columnwidth]{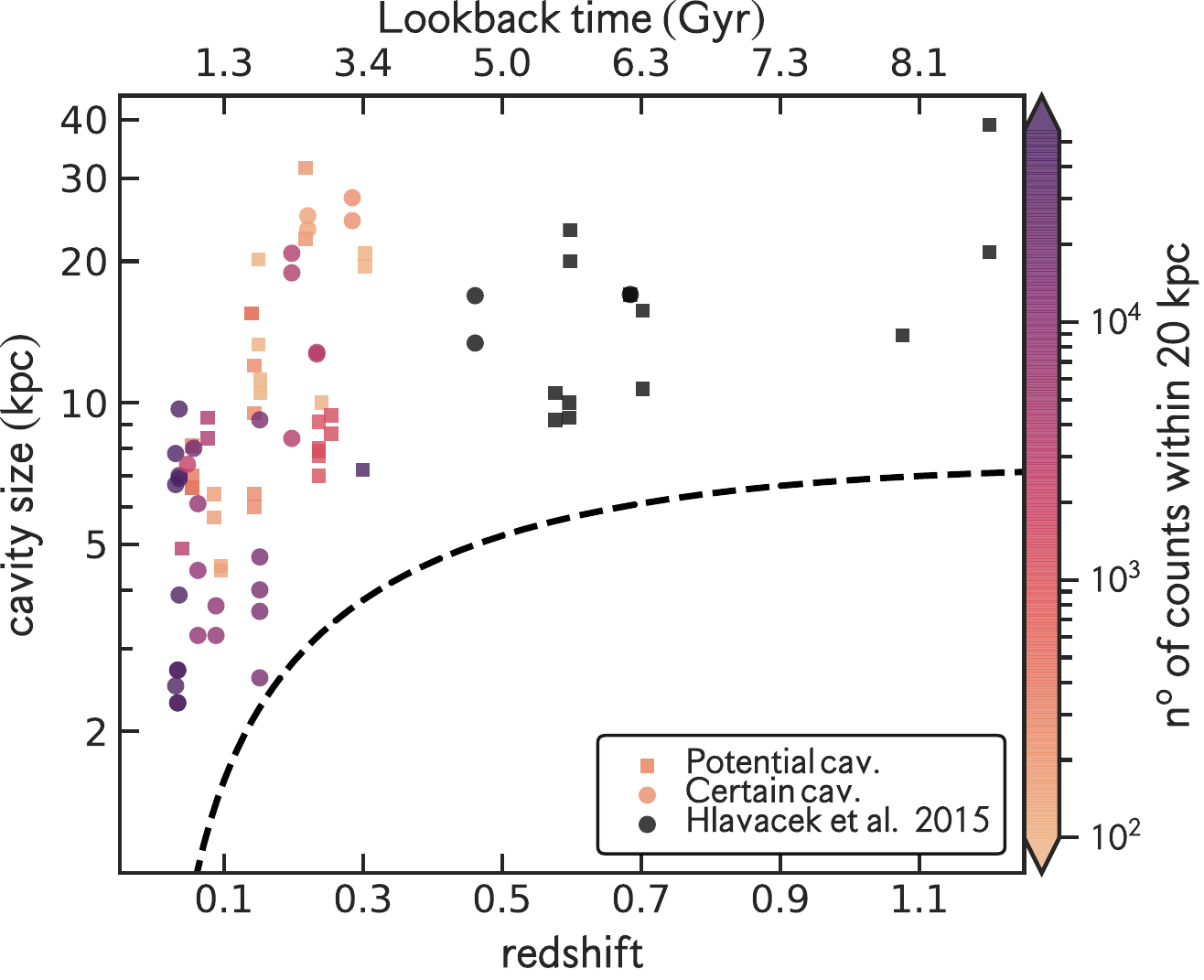}
    \caption{
    Left panel: Detection fraction of cavities as a function of redshift. The detection fraction for the entire sample is shown with a green square. The detection fraction corrected by resolution to match the SPT-SZ sample is shown with a green circle. The SPT-SZ detection fraction is displayed with an orange circle \citepalias{Hlavacek-Larrondo15}. The grey arrows correspond to the detection fraction when only ``certain'' cavities are taken into account.
    Right panel: Cavity size versus redshift color-coded by the number of counts within the central 20~kpc. We have also included cavity measurements from the high-$z$ SPT–SZ sample \citepalias{Hlavacek-Larrondo15}, shown with black symbols.
    Certain (C) cavities are displayed with circles, whereas potential (P) cavities with square symbols. 
    The dashed black line corresponds to two times the size of the \textit{Chandra} PSF as a function of redshift. In both panels, the upper axis gives the lookback times in Gyr.  \label{fig:smaller_cav_size}}
\end{figure*}

\section{Results and Discussion}\label{sec:results}
\subsection{Detection fraction of cavities and evolution}\label{sec:fraction_cav}

Overall, we detected 67 X-ray cavities in 30 clusters, of which 32 are classified as certain (C) and 35 as potential (P) cavities. From the CC cluster sub-sample, we find 29 clusters with cavities, of which 12 clusters reveal mostly certain cavities and 17 potential cavities. The remaining 34 CC clusters lack X-ray depressions. We also find in one NCC cluster, G269.51+26.42, two potential cavities located in opposite sides of the cluster center. The rest of the NCC clusters show no hint of X-ray depressions. We find that most of the detected cavities come in pairs, and they are usually located on opposite sides of the cluster core, as expected, considering that X-ray cavities are believed to be inflated by radio jets. It is worth mentioning that 29/30 of clusters with cavities also have radio emission associated with the central source (Olivares et al. in prep). Some clusters show multiple X-ray cavities, likely due to either multiple AGN outbursts or the disintegration of large cavities, while five clusters have single cavities \citep[e.g.,][]{morsony10,Canning_2013}. 

In total, 18\% of all clusters in our sample, including both CC and NCC clusters contain X-ray cavities (see Fig~\ref{fig:smaller_cav_size}, left panel), a few times smaller than the fractions found by previous studies of nearby clusters and about twice as high as that of the high-$z$ SPT–SZ sample (7\%--9\%; \citetalias{Hlavacek-Larrondo15}). We have included uncertainties associated with the fraction of clusters with cavities using the Wilson interval method \citep{brown01}. We note that previous studies of nearby clusters tend to be biased towards X-ray bright clusters. Furthermore, our findings suggest a slightly lower duty cycle of $\sim$46\%, as 28 of the 63 CC clusters ($n_{\rm core}$>1.5$\times$10$^{-2}$~cm$^{-3}$) have detected cavities (see Fig.~\ref{fig:smaller_cav_size}, left panel), compared to previous studies which predict AGN feedback duty cycle to be high (60--90\%, \citealt[][]{birzan12,fabian12,panagoulia14b}). We stress, however, that different definitions have been used to classify cool core clusters. At high-$z$, \citetalias{Hlavacek-Larrondo15} found a lower limit of $\sim$11\% on the duty cycle for the SPT-SZ sample, as only 6 of the 52 clusters with signs of cooling reveal cavities.
 
To explore the evolution of the detection fraction of cavities, we compare our results with those found in the high-$z$ SPT-SZ sample \citepalias{Hlavacek-Larrondo15}. Bear in mind that the SPT–SZ sample is limited by resolution (see Fig~\ref{fig:smaller_cav_size}, right panel), with the smaller cavities detected in this sample having sizes of $\lesssim$10~kpc. The latter is due to a combination of larger Chandra PSF at high~$z$ and lower number of counts compared to low-$z$ clusters. To account for that limitation, we compute the detection fraction taking only clusters with cavities sizes larger than $\gtrsim$10~kpc to match the observing bias of the SPT–SZ sample. That yields a detection fraction of 9\%, which is in good agreement with the SPT–SZ sample \citepalias{Hlavacek-Larrondo15}. In the same vein, if we consider only clusters with ``certain'' (C) cavities and sizes $\gtrsim$10~kpc, the detection fraction of the Planck sample drops to 3\%, close to the 2\% obtained in the high-$z$ SPT–SZ sample when only clearly detected cavities are taken into account. These findings suggest that the AGN feedback duty cycle has remained constant over almost 8 Gyrs. This trend strongly agrees with the lack of evolution in the fraction of cool-cores clusters (of $\sim$40--60\% across the same redshift range \citealt{ruppin21}), which is linked to the ICM cooling. An absence of evolution on the detection fraction of cavities could imply that the mechanical feedback in CC clusters has been in place and maintained across almost 8 Gyrs. 

All the above is quite intriguing given that the AGN-hosting BCG fraction in the SPT-SZ cluster sample, selected from infrared WISE observations, appears to be strongly evolving with redshift (see \citealt{somboonpanyakul22,birzan17,Hlavacek_Larrondo_2013}; also \citealt[][]{silverman09,haggard10} for related studies). The authors argue that nearby clusters may grow by dry mergers without increasing the AGN activity. Whereas, high-$z$ clusters may accrete cold gas from gas-rich mergers and satellite interactions, which could drive a massive inflow of cold gas towards the central region, increasing the AGN activity. With more fuel available at high-$z$, the accretion rate is more likely to reach the Eddington limit, leading to a transition from a mechanical feedback state to a radiative feedback mode \citep[e.g.,][]{churazov05,dubois12,russell13}. Therefore, the lack of evolution on the cavity fraction at high-$z$ may be due to the dominance of BCGs with radiatively efficient AGNs (see also \citealt[][]{Hlavacek_Larrondo_2013}).

\subsection{Cooling luminosity versus Cavity power}
\begin{figure}
\centering
    \includegraphics[width=0.4\textwidth]{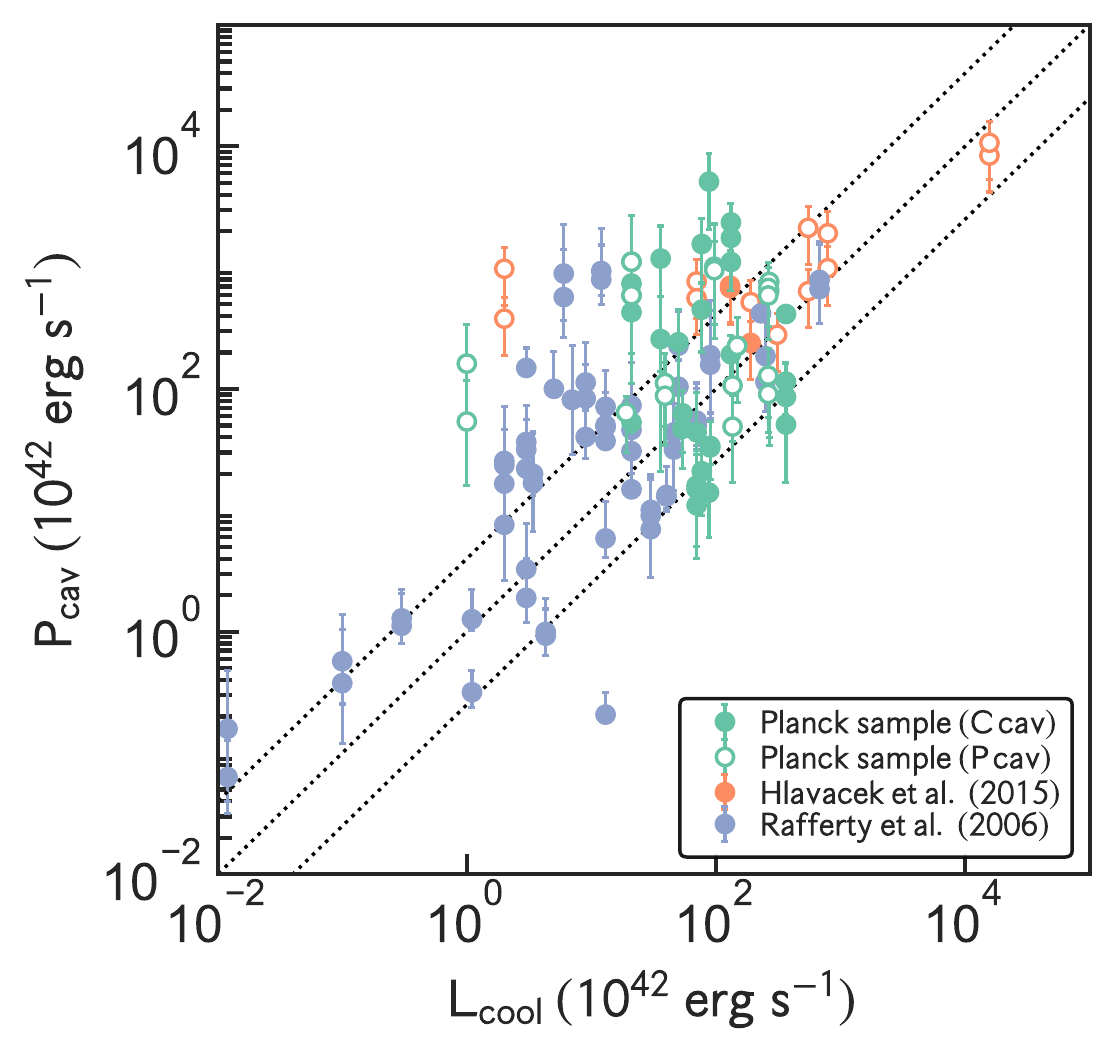}
    \caption{Comparison between the mechanical power being injected by the AGN in the BCG ($P_{\rm cav}$) and the cooling luminosity ($L_{\rm cool}$) of the cluster at 7.7~Gyrs. The dotted lines are from bottom to top are pV, 4pV, 16pV, per cavity, respectively. Certain (C) cavities are shown with filled green circles, whereas Potential (P) cavities with open green circles for Planck selected clusters. We included X-ray cavities from the high-$z$ SPT–SZ sample \citepalias{Hlavacek-Larrondo15} with filled and open orange circles for the certain and potential cavities, respectively. We have also added nearby clusters with X-ray cavities from \citet{rafferty06} sample shown with purple circles.} \label{fig:lcool_pcav}
\end{figure}

One goal of this work is to test whether the AGN is able to compensate for the cooling losses of the ICM by heating caused from radio jets. We used the cavity power ($P_{\rm cav}$) as a proxy for the mechanical power released by the AGN. The $P_{\rm cav}$ was estimated by dividing the total enthalpy of each cavity ($E_{\rm cav} = 4 p V$) by its age. Here $p$ is the thermal pressure of the ICM at the projected location of the cavity, defined as the center of each ellipse, and $V$ is the cavity volume. We assumed that the cavities have a prolate shape. The age of the cavity can be given by the buoyant rise time, the refill time, or the sound crossing time \citep{mcnamara00,birzan04,McNamara_2005}. For the purpose of this work, we used the buoyant rise time ($t_{\rm buoy}$) as the age of the cavity, as done in previous studies. The $t_{\rm buoy}$ corresponds to time for the cavity to rise buoyantly at its terminal velocity, and is defined as $t_{\rm buoy} = R / v_{\rm t}= R \sqrt{S C/2 g V}$, where $S$ is the cross-section of the bubble ($S=\pi r_{\rm b}^{2}$), $C$ (= 0.75) is the drag coefficient \citep{churazov01}. Lastly, the local gravitational potential, $g$, was derived assuming hydrostatic equilibrium.

We drew on top of each identified X-ray cavity an ellipse model, as done in previous works \citep[e.g.,][]{dong10,shin16}. Accordingly, the volume of the cavities is $V=4\pi r_{\rm b}^{2}r_{\rm a}/3$, where $r_{\rm a}$ is the semi-major axis, $r_{\rm b}$ is the semi-minor axis of each X-ray cavity (see Table~\ref{tab:cav}). We also include in Table~\ref{tab:cav} the significance of the detection for each cavity. {The significance was calculate as the surface brightness ratio of the surrounding ``blackground'', measured within the same aperture size as the cavity , and the cavity. Certain cavities have average significance of 2.2, while potential cavities have average significance of 1.5.} 


\setlength{\tabcolsep}{1pt}
\renewcommand{\arraystretch}{0.9}

\begin{table}
\caption{Cavity properties\label{tab:cav}}

\resizebox{1.0\columnwidth}{!}{
\hspace{-9.0pt}\begin{tabular}{lccccccccc}
\hline
\small {Cluster name} & {Class} & 
{$r_{\rm a}$} &
{$r_{\rm b}$} &
{R} &
{PA}  &
{$t_{\rm bouy}$} &
{$P_{\rm cav}$} &
{$L_{\rm cool}$} &
{cav. } \\
{} & {} & 
{(kpc)} &
{(kpc)} &
{(kpc)} &
{(deg.)}  & 
{( 10$^{7}$~yr)} & 
{( 10$^{44}$~erg~s$^{-1}$)} &
{(10$^{44}$~erg~s$^{-1}$)} &
{significance}\\
\hline\hline
 G021.09+33.25&C&4.1&2.6&6&140&0.9$^{+0.6} _{-3.3}$&0.9$^{+1.6} _{-0.2}$&10.1 & 2.2\\
 G021.09+33.25&C&5.4&3.6&7&70&1.6$^{+2.2} _{-1.3}$&1.2$^{+1.1} _{-1.0}$&10.1 & 2.7 \\
\hline\\
\end{tabular}}
\footnotesize{(This table is available in its entirety in machine-readable form.)}
\end{table}

Motivated from the previous studies \citep[e.g.,][]{rafferty06}, we calculate the X-ray cooling luminosity, $L_{cool}$, within a volume where the deprojected (isobaric) cooling time, $t_{\rm cool}$, is 7.7~Gyrs (see Table~\ref{tab:cav}). It is representative of the epoch of the last major. 
Since then, clusters have been relaxed and a cooling flow could develop \citep{rafferty06}. For the cooling luminosity, we used $L_{\rm cool} = \int n_{\rm e} n_{\rm H} \Lambda(T,Z) dV$, where $\Lambda(Z,T)$ is the cooling function which depends on the temperature, $T$, and metallicity, $Z$, of the hot gas. We use the cooling functions from \citet{OGnat07}, assuming metallicity of $Z=1~Z\odot$, since typical CC clusters have solar or nearly solar metallicity within their cores \citep[e.g.,][]{molendi01,mernier16}. 

In Figure~\ref{fig:lcool_pcav} we compare the mechanical power released by the AGN located in the central BCG and the cooling luminosity ($L_{\rm cool}$) of each cluster. As a matter of comparison, we included galaxy groups and elliptical galaxies from \citep{rafferty06}, as well as high-$z$ clusters from the SPT–SZ \citepalias{Hlavacek-Larrondo15}.
Notably, our Planck sample has systematically lower mechanical powers than the high-$z$ SPT–SZ sample, indicating our ability to detect smaller cavities. The smallest X-ray cavities detected in the SPT–SZ sample have sizes on the order of $\sim$10~kpc, whereas the resolution for the Planck clusters, as they are at lower $z$ clusters, is on the order of $\sim$2.5~kpc. The scatter on the $L_{\rm cool}$ versus $P_{\rm cav}$ relation is slightly smaller in our sample than in the high-$z$ SPT–SZ sample by $\sim$20\%. To make a fair comparison with the high-$z$ SPT-SZ sample taking only cavities with sizes $\gtrsim$10~kpcs, we find that the scatter is 65\% smaller compared to the SPT-SZ sample. The higher scatter on this relation for high-$z$ clusters is consistent with being fueled mainly through wet mergers \citep{somboonpanyakul22}.

{As shown in Fig~\ref{fig:lcool_pcav} the $L_{\rm cool}$ and $P_{\rm cav}$ realized from the jets are positively correlated, indicating that the energy realized by the AGN is sufficient to balance the radiative losses of the ICM within the cooling radius for most of the sources in the sample.} Some of those objects may require additional heating from another mechanism, such as thermal conduction and shocks \citep[e.g.,][]{pope05}. However, we expect that some of these clusters may be in a cooling phase discarding the need for an additional heating source. The nature of the AGN feedback is cyclic and does not always require a balance between the cooling luminosity and the AGN heating \citep{mcnamara07}. This is likely displayed on the scatter of the $L_{\rm cool}$ versus $P_{\rm cav}$ relation. In that sense, the AGN power is variable, and the objects change their $P_{\rm cav}$ depending on what phase of the AGN feedback cycle they are observed in. Another source of scatter comes from dynamically disturbed clusters due to either sloshing motions or mergers. These mechanisms move the hot gas out of the central BCG to large distances producing lower $L_{\rm cool}$ for a given $P_{\rm cav}$ value, as found for clusters with higher centroid shifts indicative of dynamically disturbed atmospheres (Olivares et al. in prep). 

\section{Limitations}\label{sec:limitations}
One of this work's limitations is the fact that we are probably missing cavities due to shallow X-ray observations, in particular in high-$z$ clusters \citep[e.g.,][]{diehl08,birzan12}. As pointed out by several studies, X-ray cavities are more easily detected in clusters with stronger cool cores, as the contrast between the depression and surroundings is sharper, and it is more difficult to find bubbles in high-redshift clusters due to the lack of counts. The detectability of cavities also decreases with their radius \citep{enblin02}.

More importantly, cavities that have sizes below the resolution (e.g., cavity size $\leq2$~kpc for clusters at $z$=0.1) will be undetectable in such an analysis. As shown in Figure~\ref{fig:smaller_cav_size}, this effect will increase at high-$z$. For example, cluster at $z$>0.5 only cavities with sizes $\geq 6$~kpc can be detected. We also note that sources with more than 2000 counts, in the central region, tend to have ``certain'' detected X-ray cavities (circle symbols). Therefore, we stress that deeper \textit{Chandra} follow-up observations are required to confirm the presence of any potential X-ray cavity, especially in high-$z$ clusters. Aside from the data quality, other effects may be also interfering with the cavity detectability, such as orientation, location, and angular size (see \citealt{enblin02, diehl08, bruggen09} for more details). {As pointed out by \citet{enblin02}, the detectability of cavities decreases with distance to the center, and for a cavity moving in the plane of the cavities, the contrast decreases slowly. Cavities that lie off the plane of the sky have reduced contrast and therefore are harder to detect. To quantify this projection effect, we assume a random distribution of the angle of the cavity relative to the plane of the sky, a typical cavity size of 10~kpc, and a typical beta profile for the ICM distribution (r$_{c}$=20, beta=3/4). At an average projected distance of 30~kpc, 20\%--30\% of the cavities would have a contrast below our detection limit and would have been missed in our study.  
}


It should be noted that the $P_{\rm cav} - L_{\rm cool}$ relation is also affected by projection effects, likely introducing scatter. As pointed out by \citet{diehl08}, all the physical quantities involving the cavity power, $P_{\rm cav}$, such as density, temperature, and pressure, are measured at the projected distance from the cavity to the center rather than the true distance. The former corresponds to a lower limit. The pressure increases towards the center, leading to an overestimation of the ambient pressure at the cavity position. On the other hand, the cavity ages will be underestimated as they are proportional to the cavity distance. Both mentioned effects will bias the cavity power upwards.

\section{Conclusions}\label{sec:conclusions}
We have investigated the mechanical AGN feedback mechanism in central cluster galaxies using archival X-ray \textit{Chandra} observations of 164 Planck selected clusters to search for X-ray cavities. 

(i) Using several techniques to look for X-ray cavities, including inspection of the original image, a model subtracted image and an unsharp masked image, we find 65 X-ray cavities in 29 systems out of 63 CC clusters. Among them, 12 systems have clearly detected cavities, whereas 17 have only potential depressions. Two potential cavities were also found in one NCC cluster.

(ii) We measured a total detection fraction of X-ray cavities of $\sim$18\%, twice the detection rate of the high-$z$ SPT–SZ sample, indicating that clusters have radio-mode feedback only 18\% of the time. Nevertheless, our detection fraction of 9\% is close to the high-$z$ SPT–SZ sample when taking only cavities with sizes $\gtrsim$10~kpc to match the resolution of the SPT-SZ sample. We interpreted this as an lack of evolution of the AGN feedback cycle across cosmic time.

(iii)  We find that the AGN heating traced by the power of the X-ray cavities alone is able to balance the radiative losses of the ICM in our sample. Our sources have slightly lower cavity power per cavity than high-$z$ massive clusters from the SPT-SZ sample due to smaller cavities being detected in our sample.

\noindent Future high-resolution X-ray observations from \textit{Chandra} satellite and the upcoming Advanced X-ray Imaging Satellite (AXIS) telescope will be needed to find more cavities in the faintest clusters and confirm the discussed findings in high-$z$ clusters.

\section*{Acknowledgments}
This research has made use of software provided by the \textit{Chandra} X-ray Center (CXC) in the application packages CIAO. V.O. and Y.S. were supported by NSF grant 2107711, Chandra X-ray Observatory grant GO1-22126X, and NASA grant 80NSSC21K0714. 

\section*{Data Availability}
The \textit{Chandra} raw data used in this paper are available to download at the HEASARC Data Archive website\footnote{https://heasarc.gsfc.nasa.gov/docs/archive.htm}.




\bibliographystyle{mnras}





\bsp	
\label{lastpage}
\end{document}